# Arc statistics with realistic cluster potentials

# II. Influence of cluster asymmetry and substructure

Matthias Bartelmann*, Matthias Steinmetz, and Achim Weiss

Max-Planck-Institut für Astrophysik, Postfach 1523, D–85740 Garching, FRG

24 August 1994

**Abstract.** We construct a sample of numerical models for clusters of galaxies and employ these to investigate their capability of imaging background sources into long arcs. Emphasis is laid on the statistics of these arcs. We study cross sections for arc length and length-to-width ratio and optical depths for these arc properties, we examine the distribution of arc widths and curvature radii among long arcs, and we compare these results to predictions based on simplified (radially symmetric) cluster models. We find that the capability of the numerically modeled clusters to produce long arcs is larger by about two orders of magnitude than that of spherically symmetric cluster models with the same observable parameters (core radii and velocity dispersions), and that they are similarly efficient as singular isothermal spheres with the same velocity-dispersion distribution. The influence of source ellipticity is also investigated; we find that the optical depth for arcs with a length-to-width ratio $\gtrsim 10$ is significantly larger for elliptical than for circular sources. Given these results, we conclude that spherically symmetric lens models for galaxy clusters, adapted to the observable parameters of these clusters, grossly underestimate the frequency of long arcs. We attribute this difference between numerically constructed and simplified analytical lens models to the abundance and the extent of intrinsic asymmetry and to substructure in galaxy clusters.

**Thesaurus codes:** 12.04.1 – 12.07.1 – 11.03.1

## 1 Introduction

The statistics of long arcs in clusters of galaxies can provide important information about the distribution of the probably dominant dark matter in clusters which is otherwise hard to extract. Unfortunately, the interpretation of arc statistics is by no means straightforward since it critically depends on the lens models chosen to parametrize the lensing cluster population. Several studies have been performed on the basis of analytical lens models (see, e.g., Miralda-Escudé 1993a,b, Wu & Hammer 1993, and references therein). Wu & Hammer arrive at the conclusion that either cluster core radii are much smaller

---

* Present address: Harvard University Observatory, 60 Garden Street, Cambridge MA 02138, USA

*Send offprint requests to:* M. Bartelmann (Cambridge address)

than observed in the galaxy distribution or cluster mass profiles are much steeper than isothermal in order to concentrate a larger fraction of the cluster mass towards the cluster center; otherwise, the total number of long arcs in clusters of galaxies would be far too low.

We argued in a previous paper (Bartelmann & Weiss 1994, henceforth Paper I), that analytical (and therefore simplified) lens models might seriously underestimate the capability of galaxy clusters to form arcs. The reason for this suspicion was that clusters are typically not well-relaxed, spherically symmetric systems, but that they are asymmetric and frequently substructured to a high degree. The shear induced by the intrinsic ellipticity of cluster components and by the presence of several mass concentrations can notably increase the capability of clusters to form arcs.

We have described in Paper I the method we employ for investigating the lensing properties of a numerically simulated galaxy cluster. We will therefore only give a brief sketch of this technique here except where it deviates from the prescription given in Paper I, and refer the reader to Paper I for further details.

The present paper describes in Sect.2 the construction of the cluster sample and some properties of the simulated clusters. The cross sections of the clusters for producing arcs with specified properties are the subject of Sect.3. In Sect.4, cross sections are integrated to yield optical depths, and Sect.5 contains a summary and a discussion of the results. Throughout, we use a Hubble constant of $H_0 = 100\,h$ km/s/Mpc, where $h = 0.5$.

## 2 The cluster sample

### 2.1 Computational requirements

Cosmological $N$-body simulations aiming at realistic cluster potentials require a careful selection of the particle number, the simulation volume, and the gravitational softening. First of all, the simulation volume should be large enough to contain a suitable number of massive clusters. Since typically only one percent of the mass of the universe can be found in clusters with a mass above $10^{15}$ M$_\odot$ (Frenk et al. 1990), the simulation volume has to have a mass of several $10^{18}$ M$_\odot$ (or equivalently a cube side length of about $150h^{-1}$ Mpc) to enclose about a dozen rich clusters. Second, the radius of a cluster region which can become critical to strong lensing is of the order $100h^{-1}$ kpc, i.e. the softening should be smaller than about $25h^{-1}$ kpc. Furthermore, such a region should contain at least a few hundred particles to avoid a contamination of the results from sampling errors and two-body relaxation effects. Since a cluster of $10^{15}$ M$_\odot$ with a roughly isothermal profile and a radius of about $1\ldots 2h^{-1}$ Mpc typically has a mass of the order of a few $10^{14}$ M$_\odot$ within the region responsible for strong lensing phenomena, the particle mass should not much exceed $10^{11}$ M$_\odot$. We can therefore conclude that one needs a dynamic range of seven orders of magnitude in mass and four orders of magnitude in length, which challenges even the current generation of supercomputers. It also becomes evident that such a resolution in length cannot be attacked by particle-mesh schemes; rather, one has to use P3M or tree codes (whose spatial resolution is not fixed by the dimensions of mesh cells). Since by definition galaxy clusters are highly overdense objects, the application of tree-based methods seems to be advantageous. Furthermore, roughly 1500 time steps are necessary to propagate the $N$-body system with sufficient accuracy from $z = 15$ (where the fluctuations on the scales considered are still linear) to $z = 0$. Note that even a $180^3$-particle simulation of a $150h^{-1}$ Mpc box, which is the resolution of most of the simulations presented below, only marginally fulfills these requirements.

## 2.2 Numerical method

Since with such a high particle number even tree codes are expensive in CPU ($> 400$ hours on a Cray YMP) and main memory ($\simeq 4$ GByte), we have used a hybrid scheme (the so-called tree pruning) as originally proposed by Porter (1985) and used in different modifications by Zurek et al. (1986) and Navarro & White (1994). The basic idea of these hybrid schemes is very similar to the principles of a tree code: a typical cluster mass of $10^{15}$ M$_\odot$ corresponds to a homogeneous sphere with a radius of $7.5h^{-1}$ Mpc. But scales larger than about $10h^{-1}$ Mpc are still linear, which means that most of the particles which belong to a cluster originally stem from a region not much larger than $10h^{-1}$ Mpc. To describe the tidal field exerted by much more distant particles, however, it is not necessary to resolve them in detail, but it is sufficient to group them together as in a standard tree code and to consider only the monopole moments of the groups. Since on scales above $10h^{-1}$ Mpc, where the density evolution can be considered linear, the particles roughly remain confined in their original volume, it is sufficient to construct the tree only once. Distant particles are grouped into nodes which thereafter are considered as macro-particles, and all substructure below such distant nodes is removed (tree pruning). Afterwards, these macro-particles are considered as normal but more massive collisionless particles. Thus, $N$ particles surrounding a potentially forming cluster can be replaced by $\propto \log N$ macro-particles. To avoid two-body relaxation effects between particles of very different mass, the gravitational softening length $\varepsilon$ of a macro-particle of mass $m$ has to be increased; we take $\varepsilon \propto m^{1/3}$. In summary, we replace one large $N$-body simulation by a small series of much smaller $N$-body simulations.

To be more specific, we start with a coarsely grained $36^3$-particle, COBE-normalized (Efstathiou et al. 1992) CDM $N$-body simulation ($\Omega = 1, h = 0.5$ as throughout). At $z = 0$, we search for regions of radius $5h^{-1}$ Mpc which contain more than 100 particles, i.e. which have a mean overdensity of more than 14. We found 13 of these regions in our simulation. All these regions were recalculated by the method described above. The particles are traced back to $z = 15$. There, they are typically confined within a nearly spherical region of about $8 \ldots 10h^{-1}$ Mpc. In order to safely describe the complete mass flow towards the cluster with low-mass particles only, we take a sphere $\Upsilon$ of radius $R_\Upsilon = 12.5h^{-1}$ Mpc as the highly resolved region. Each particle within $\Upsilon$ is replaced by 125 low-mass particles, and small-scale power is added according to the new Nyquist frequency. The gravitational softening for particles of the lowest mass is chosen to be $\varepsilon = 25h^{-1}$ kpc. Particles outside the sphere are grouped to nodes of size $d$ according to the tree algorithm. Those nodes are pruned (i.e. replaced by a macro-particle), for which the ratio of size $d$ to the distance to surface of $\Upsilon$, given by $r - R_\Upsilon - d$, is larger than a given error tolerance $\theta$; $r$ being the center-of-mass distance between the node and $\Upsilon$ (for more details on tree construction, tree pruning etc. see, e.g., Porter 1985, Steinmetz & Müller 1993). Test simulations with this kind of boundary conditions have shown that the results can be further improved if the tree is constructed at a centered time point, e.g. at $z = 1$, which is the time center if the cosmic scale factor is used as time coordinate. The time-centered particle positions can simply be obtained by using the corresponding particle distribution of the coarsely grained presimulation. Another possibility is to propagate the particles according to the truncated Zel'dovich approximation with a truncation length of a few Mpc. A typical high-resolution run consists of about 75000 low mass and 8000 macro-particles. The reduction procedure requires about 5 hours per model, the following $N$-body simulation about 10 hours, where all timings are for one processor of a CRAY YMP. The memory requirement is less than 100 MByte. We point out that the reduction method is by no means optimized; we expect that timings below 1 hour could relatively easily be obtained. Furthermore note that the macro-particles are already preprocessed by a tree code. Therefore, the interaction list of macro-particles is relatively long; it is comparable with the neighbour list of elementary particles (a few hundred). Therefore, the CPU cost for such a simulation is substantially higher than for a simulation with the same total particle number but without macro-particles. A detailed investigation of the accuracy of this simulation technique in the context of cluster simulations is postponed to a forthcoming paper.

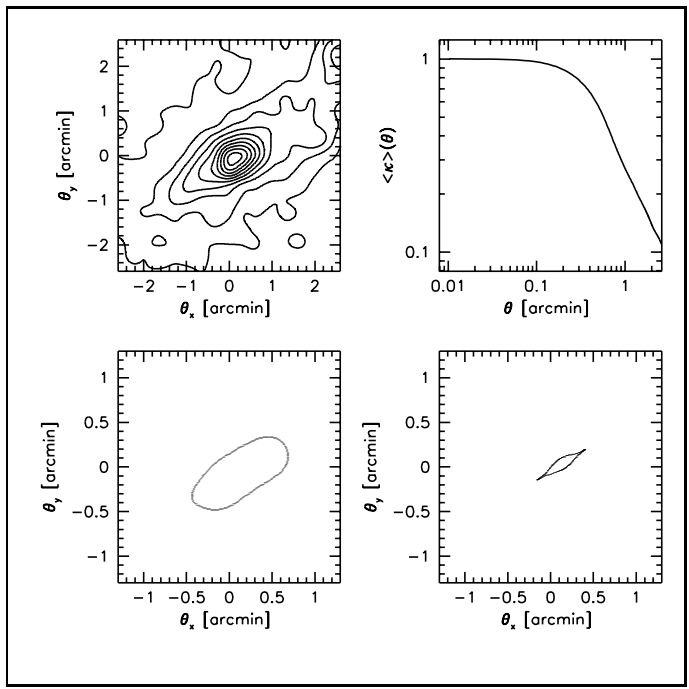

**Fig. 1.** Example of one cluster model at $z = 0.35$. Upper left frame: convergence $\kappa$, i.e., surface mass density scaled by its critical value; upper right frame: azimuthally averaged $\kappa$ profile; lower left frame: critical curve; lower right frame: caustic. Note the change of scale in the lower frames to enlarge the critical curve and caustic. See also Fig.3 for arcs produced by this cluster

### 2.3 Cluster properties

At $z = 0$, the 13 clusters typically contain a dominant, but not fully relaxed clump. The density stratification of these main objects is well described by a power law with index $n \simeq -2.3$ and a core radius of about 50 kpc/$h$. Superposed on this density profile, there exist smaller subclumps with a typical mass of about $10^{13}$ M$_\odot$; cf. Fig.1 for an example of the projected surface-mass density of one cluster model at $z = 0.35$, together with its azimuthally averaged profile, its critical curve, and its caustic.

Except one cluster (700 km/sec), the typical velocity dispersion ranges between 1000 and 1500 km/sec. In most of the clusters, we find 2...4 accompanying secondary clumps with a mass ranging between 5...30% of the main object's mass. However, as we will see later on, the mass of these secondary clusters is too small to become critical to strong lensing. Tracing back the history of the main clusters we found that none of these clusters has a dominant progenitor at $z \gtrsim 0.3$, but is typically formed out of a dozen small mass progenitors as is typical for clusters formed in a high-$\Omega$ hierarchical

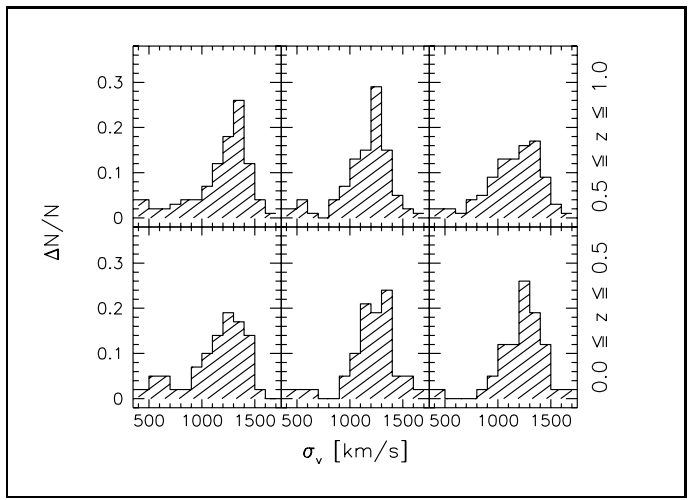

**Fig. 2a.** Histograms of the line-of-sight velocity dispersion $\sigma_v$ of cluster models, in km/s. The top frames show the velocity-dispersion distribution for cluster models with redshifts $0.5 \leq z < 1.0$, while the bottom frames are for cluster models with $0 \leq z < 0.5$. The three panels per row are for three orthogonal directions of projection. It is seen that (1) the velocity-dispersion distribution is isotropic, as expected, and (2) that the average velocity dispersion in the low-redshift subsample is slightly larger than in the high-$z$ subsample. The histograms are normalized by the total number of models (see Sect.3.1 for details)

universe. For later use we point out that in the redshift interval $z = 0 \ldots 0.5$, which is of greatest importance for lensing phenomena, all the mass which belongs to the cluster at $z = 0$ can be found within a sphere of radius $5h^{-1}$ Mpc. This statement approximately still holds if we extend the corresponding interval to $z \in [0 \ldots 1]$.

To check whether the cluster models are suitable as potential wells for arc statistics, we have performed a bootstrap analysis (for details see Heyl et al. 1994 and references therein) to see whether the potentials are significantly influenced by individual particles. The main idea can be outlined as follows. Consider an ensemble of $N$ particles resulting from an $N$-body simulation. From that ensemble, one creates a series of new ensembles by drawing $N$ integer random numbers equally distributed in the interval $[1, N]$. These $N$ random numbers are used as indices for all those particles included into the new ensemble; i.e. some particles are taken more than once into the new ensemble, and some particles are left out. In the statistical mean $1/e \simeq 37\%$ of the particles of the original ensemble do not enter the bootstrapped sample. A feature (e.g., a subclump) which only occurs in a few bootstrapped ensembles is, therefore, likely to be caused by only a few particles, whereas features which occur in all ensembles can be considered statistically significant. By comparing different ensembles, one can get an error estimate for properties derived from the particle distribution; e.g., the cross section of a cluster for creating large arcs. For more details about statistical bootstrapping we refer to Heyl et al. (1994) and to Efron (1982) and Efron & Tibshirani (1986).

Figures 2a and 2b serve to summarize such properties of the cluster models which are in principle observable, namely their velocity dispersion and core radius distributions. In both these figures, the top frames are for high, the bottom frames for low redshifts. The mean velocity dispersion distribution is $\langle \sigma_v \rangle \simeq 1100 \pm 250$ km/s independent of the cluster redshift. For the core radii, we obtain $\langle r_c \rangle \simeq 80 \pm 20$ kpc/$h$ for high and $\langle r_c \rangle \simeq 50 \pm 10$ kpc/$h$ for low redshifts; this decrease is caused by merging of substructures which reduces the azimuthally averaged core radii.

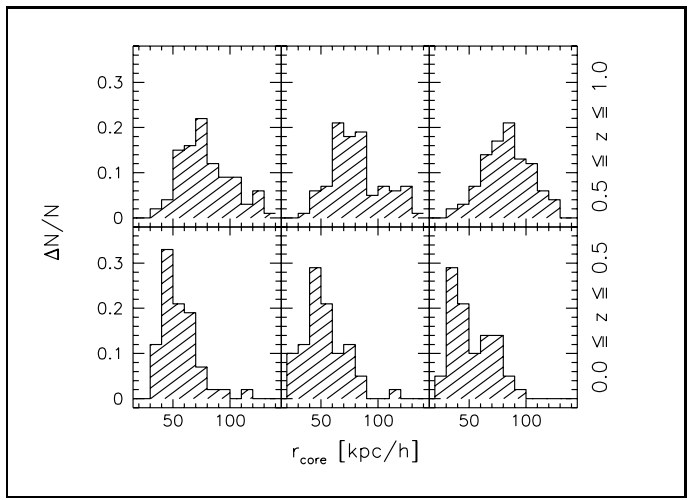

**Fig. 2b.** Histograms of cluster core radii, organized analogous to the histograms in Fig.2a. The core-radius distribution is also isotropic. It becomes narrower and peaks at smaller core radii in the low-$z$ subsample

## 3 Cross sections

### 3.1 Numerical cluster models

The lensing cross section $\sigma(Q)$ for images with property $Q$ is defined to be the area in the source plane within which the source has to lie in order to be imaged with property $Q$. We will consider here cross sections for four different arc properties $Q_l$, namely

$$
\begin{aligned}
Q_{\mathrm{L}} &: \quad \text{arc length } L \geq L_0 \ ; \\
Q_{\mathrm{W}} &: \quad \text{arc width } W \geq W_0 \ ; \\
Q_{\mathrm{R}} &: \quad \text{arc curvature radius } R \geq R_0 \ ; \\
Q_{(\mathrm{L/W})} &: \quad \text{arc length-to-width ratio } (L/W) \geq (L/W)_0 \ .
\end{aligned}
\tag{3.1}
$$

For simplicity, we denote these properties as superscripts on $\sigma$, e.g., $\sigma^{(\mathrm{L})}$ denotes the cross section for arc length $L \geq L_0$, where $L_0$ remains to be chosen. If we refer to cross sections for any property $Q_l$ without further specifying $Q_l$, we write $\sigma^{(l)}$.

To determine these cross sections, we image sources on the source plane, which is assumed to be at redshift $z_\mathrm{s} = 1$ throughout, using the deflection angles determined from the numerical cluster model. As described in detail in Paper I, we choose the source positions on the source plane such that the sources are placed on grids whose resolution is refined towards caustics, and the sources are assigned statistical weights in proportion to the inverse of the squared grid resolution. The images are then classified using the algorithm described in Paper I. For each image, we determine length, width, and curvature radius. This is done for each time step and each projection direction of each cluster model, if the cluster model is critical, i.e., if it produces critical curves, and if the cluster model has redshift $z_\mathrm{d} < z_\mathrm{s}$. In total, we have classified 73395 images in that way, which should provide a reasonably large sample to treat them statistically. As an example, Fig.3 shows arcs produced by the cluster model whose surface-mass density contours are displayed in Fig.1. The sources for this figure were randomly distributed such that the spatial image density in the *lens* plane is constant; this is to account for the magnification bias on a sample of sources whose luminosity distribution has a power-law index close to $-1$. A source distribution with constant number density in the *source* plane would be diluted behind the cluster because of the solid-angle distortion due to gravitational light deflection, but for the given slope of the luminosity distribution, this dilution is exactly cancelled by the magnification which allows for fainter sources to be seen behind the cluster.

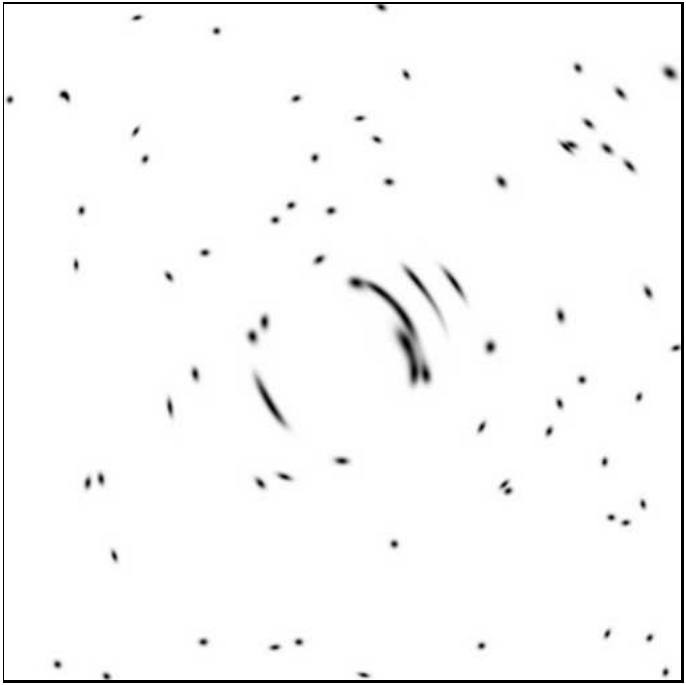

**Fig. 3.** Examples of arcs produced by the cluster displayed in Fig.1

The sources are chosen either circular or elliptical, where the circular sources have a diameter of $1''$. The axis ratio (small divided by large semi-axis) for elliptical sources is drawn randomly from the interval $[0.5, 1]$, and their area is chosen equal to that of the circular sources, i.e., $\pi$ square arc seconds. The surface-brightness distribution over the sources is taken to be constant for the determination of the arc properties, and Gaussian for the images displayed in Fig.3.

Next, from the number of occurrences of images with property $Q_l$, and from their statistical weights, we determine the fraction of the area covered with source positions within which images with property $Q_l$ can occur, and multiply this (so far dimensionless) area with the squared length scale in the source plane. Since we study only the central sixteenth of the lens plane for each cluster (cf. Paper I for details), and since the comoving side length of the field containing the cluster is 5 Mpc/$h$, the length scale in the lens plane is

$$\xi_0 = \frac{5}{(1 + z_{\rm d})} \frac{1}{4} \text{ Mpc}/h , \qquad (3.2)$$

and the corresponding length scale in the source plane is

$$\eta_0 = \frac{D_{\rm s}}{D_{\rm d}} \xi_0 , \qquad (3.3)$$

where $D_{\rm d,s}$ are the angular-diameter distances to the lens and source planes, respectively. Thus, we end up with a table of cross sections $\sigma_{ijk}^{(l)}$ in units of $(\text{Mpc}/h)^2$, for cluster model $i$, projected along direction $j$, at timestep $k$, for arc property $Q_l$.

When the cluster model is at low redshift, it covers a larger solid angle on the lens plane than at high redshift. To guarantee a given angular resolution of the images, the grid on which the deflection angle is defined therefore has to be refined when the cluster model approaches the observer. We start with a resolution of $512^2$ grid points at high redshift and refine this grid by bilinear interpolation such that the size of a grid cell does not exceed $0.5''$.

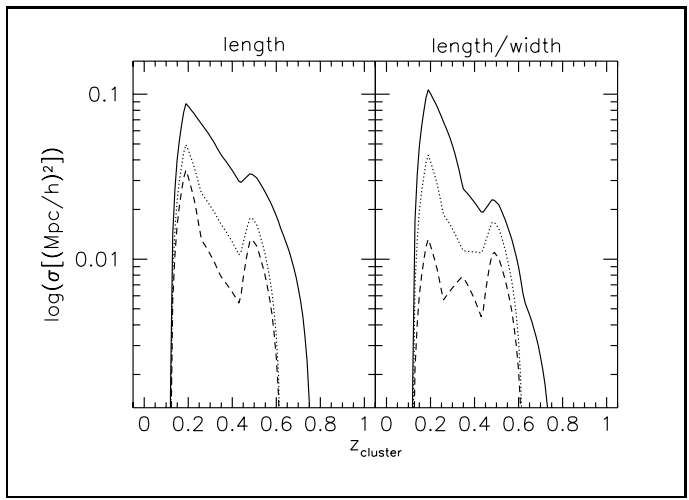

**Fig. 4.** Cross sections $\sigma_{3,2}^{(L)}(z)$ (left frame) and $\sigma_{3,2}^{(L/W)}(z)$ (right frame); i.e., these cross sections were obtained from cluster model 3 projected along the 2-axis, and they are for arc length $L \geq L_0$, where $L_0 \in \{5, 7.5, 10\}''$ (solid, dotted, and dashed curve, repectively), and for arc length-to-width ratio $(L/W) \geq (L/W)_0$, with $(L/W)_0 \in \{5, 7.5, 10\}$ in the same order of line types

The timesteps of any two cluster models do usually not coincide, and for a single cluster model, they are not evenly spaced in either redshift or time because the choice of the timesteps depends on the particle configuration in the cluster model. The next step is therefore to convert discrete tables of cross sections $\sigma_{ijk}^{(l)}$ at irregularly spaced redshifts into smooth functions $\sigma_{ij}^{(l)}(z)$, where the indices $i,j$ again label the cluster model and projection direction, respectively. This is done by second-order polynomial interpolation. As an example, we show in Fig.4 the functions $\sigma_{3,2}^{(L)}(z)$ and $\sigma_{3,2}^{(L/W)}(z)$, i.e., for cluster model 3, projected along the 2-axis, for arc length $L \geq L_0$ and arc length-to-width ratio $(L/W) \geq (L/W)_0$.

Fig.4 shows that the cross section as a function of cluster redshift can have several local maxima and minima. This is caused by merging of substructures, which increases the surface mass density if the merging direction does not conincide with the line-of-sight, and by rotation of the cluster, which can either increase or decrease the surface mass density. In addition, the cross section depends on the critical surface mass density, which is a function of distance. For a lens close to the observer or the source, the critical surface mass density (see, e.g., Eqs.5, 6 and 10 of Paper I) becomes large, rendering the cross section very small.

For each direction of projection and for each arc property, we now average the cross sections over all cluster models,

$$\langle \sigma \rangle_j^{(l)}(z) = \frac{1}{13} \sum_{i=1}^{13} \sigma_{ij}^{(l)}(z) \,. \tag{3.4}$$

The result is displayed for $\langle \sigma \rangle_j^{(L)}(z)$ and $\langle \sigma \rangle_j^{(L/W)}(z)$, for elliptical sources in Fig.5a and for circular sources in Fig.5b.

The frames in the three rows in Figs.5a and 5b show the averaged cross sections for the three projection directions. Apart from minor differences, the curves in these frames are very similar, showing that the cross sections are isotropic in the sense that none of the projection directions is in any sense preferred. Although this is to be expected, it is reassuring to see that the spatial orientation of the cluster models is not aligned with the orientation of the larger simulation volume where they were taken from.

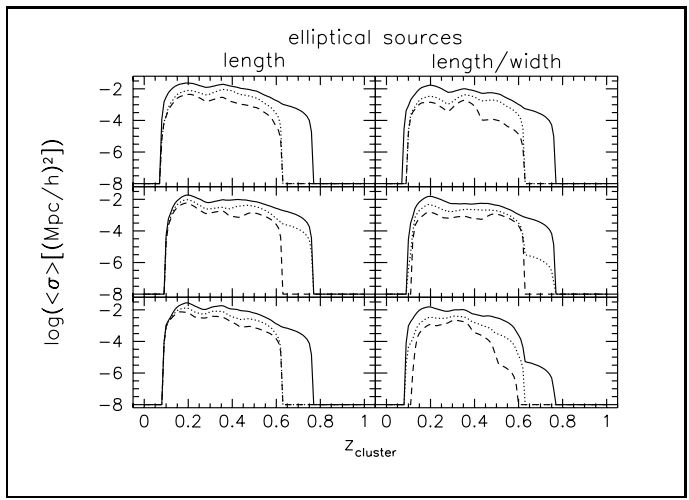

**Fig. 5a.** Averaged cross sections for long arcs from elliptical sources as functions of the cluster redshift. The solid (dotted, dashed) lines are for arc lengths $\geq 5$ (7.5, 10)$''$, respectively, or length-to width ratios $\geq 5$ (7.5,10). The three panels per column show the averaged cross sections for the three orthogonal directions of projection

For intermediate redshifts, the cross sections for arcs with $L \geq 10''$ are about an order of magnitude lower than for $L \geq 5''$, and the redshift range where long arcs can be formed is slightly narrower than for arcs of intermediate length. This leads to the expectation that, the more 'spectacular' the arcs are, the lower the scatter in the lens redshift should be. A comparison between the curves in the two columns of Figs.5a and 5b shows that the cross section for long arcs is slightly larger than for arcs with large length-to-width ratio. This means that most long arcs are also magnified in the radial direction. Comparing Figs.5a and 5b, we see that the cross sections for long arcs are insensitive to whether the sources are elliptical or circular, but that arcs with a large length-to-width ratio are more likely to occur if the sources are elliptical. This indicates that searching for long *and* thin arcs prefers such sources which are already intrinsically elliptic. We have also investigated whether long and thin arcs might preferentially be formed from such elliptical sources whose position angle is aligned with the position angle of the shear field of the cluster at the location of the arcs. There is no clear evidence for this to be true; the absolute difference between the position angles of the source and of the shear field is scattered in the interval $[0, \pi]$ without showing a significant correlation with arc length or length-to-width ratio.

Now that we have seen that the cross sections are independent of the direction of projection, we average over the three directions to obtain

$$\langle \sigma \rangle^{(l)}(z) = \frac{1}{3} \sum_{j=1}^{3} \langle \sigma \rangle_j^{(l)}(z) \,. \tag{3.5}$$

We display the cross sections $\langle \sigma \rangle^{(L)}(z)$ and $\langle \sigma \rangle^{(L/W)}(z)$ for elliptical and circular sources in Fig.6.

Fig.6 summarizes what has been discussed above: arcs with large length-to-width ratio are less frequent than long arcs, and long and thin arcs are easier to produce if the sources are elliptic instead of circular. As mentioned in Sect.2, we have checked by a bootstrapping analysis whether the surface-mass density of the cluster models is influenced by few individual particles. This is not the case; the cross sections change by at most a few percent within the bootstrapped cluster samples.

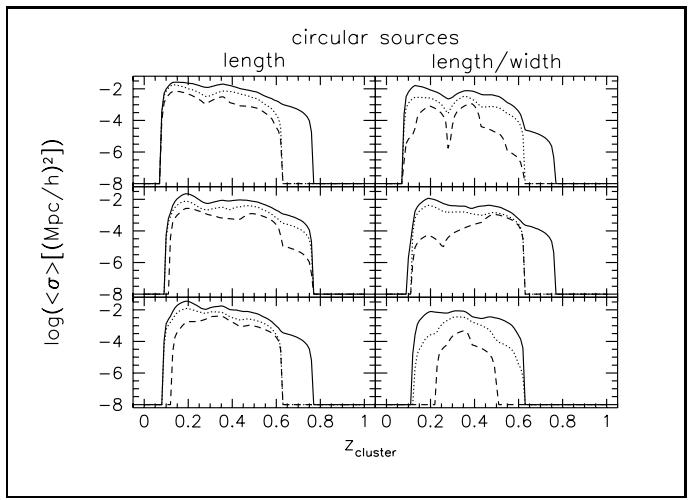

**Fig. 5b.** Cross sections analogous to those of Fig.5a, but from circular sources. A comparison with Fig.5a shows that the cross sections for *long* arcs (left column) are basically identical to those for elliptical sources, but the cross sections for *long and thin* arcs are smaller than for elliptical sources. See also Fig.6

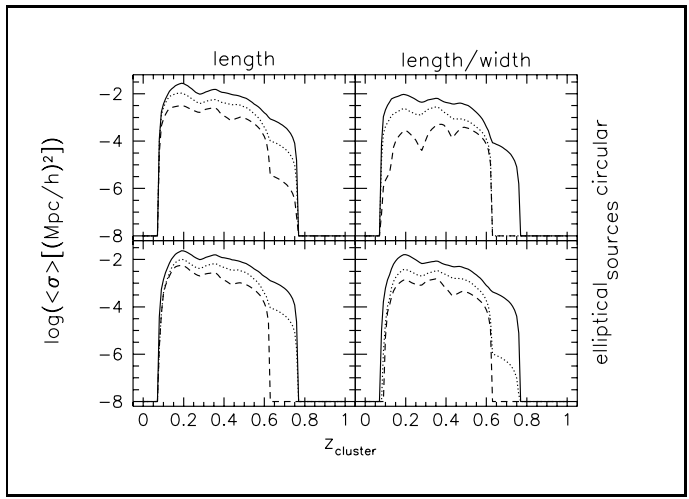

**Fig. 6.** The cross sections of Figs.5a and 5b, averaged over the three projection directions. The top row shows cross sections for circular, the bottom row for elliptical sources. The figure illustrates that elliptical sources are more likely to be imaged as long *and* thin arcs than are circular sources, although this effect is weak

### 3.2 Analytic approximations

It is instructive to compare the results obtained from the numerically modelled clusters with analytic approximations to these models. We investigate two analytic models, the singular and the non-singular isothermal sphere.

**3.2.1 Singular isothermal spheres.** The lens equation for the singular isothermal sphere reads

$$y = x - \frac{x}{|x|} \, , \tag{3.6}$$

where $y, x$ are dimensionless distances of the source and its images from the intersection points of the optical axis with the source and the lens plane, respectively, where the optical axis connects the observer's position with the center of the lens. The length scales in the source and the lens planes are $\bar{\eta}_0$ and $\bar{\xi}_0$, respectively, with

$$\bar{\xi}_0 = 4\pi \left(\frac{\sigma_{\rm v}}{c}\right)^2 \frac{D_{\rm d} D_{\rm ds}}{D_{\rm s}} \, , \; \bar{\eta}_0 = \frac{D_{\rm s}}{D_{\rm d}} \bar{\xi}_0 = 4\pi \left(\frac{\sigma_{\rm v}}{c}\right)^2 D_{\rm ds} \, , \tag{3.7}$$

where $D_{\mathrm{ds}}$ is the angular-diameter distance between the lens and the source. For $y < 1$, two images occur at positions $x_{1,2} = y \pm 1$. The moduli of the eigenvalues of the Jacobian matrix of the lens mapping are

$$\lambda_{\mathrm{r}} = 1 \ , \ \lambda_{\mathrm{t}} = \left| 1 - \frac{1}{x} \right| \ , \tag{3.8}$$

and therefore the length-to-width ratio of arcs from circular sources can be approximated by

$$\frac{L}{W} \simeq \frac{\lambda_{\mathrm{r}}}{\lambda_{\mathrm{t}}} = \frac{1}{\left|1 - \frac{1}{x}\right|} \ . \tag{3.9}$$

For $(L/W) \geq (L/W)_0$, the absolute arc position has to fulfil one of the following inequalities,

$$\begin{aligned} 1 < |x| &\leq \frac{(L/W)_0}{(L/W)_0 - 1} \ , \\ 1 > |x| &\geq \frac{(L/W)_0}{(L/W)_0 + 1} \ . \end{aligned} \tag{3.10}$$

It follows from Eq.(3.10) and the lens equation (3.6) that for interesting length-to-width ratios, e.g., $(L/W) \geq 5$, $y < 1$, and therefore two arcs occur simultaneously at positions $x_{1,2}$. For

$$\frac{1}{(L/W)_0 + 1} < y \leq \frac{1}{(L/W)_0 - 1} \ , \tag{3.11a}$$

only the outer arc (at $x_1$) fulfils $(L/W) \geq (L/W)_0$, while for

$$y \leq \frac{1}{(L/W)_0 + 1} \ , \tag{3.11b}$$

both arcs fulfil the length-to-width criterion. Thus, the dimensionless cross section for $(L/W) \geq (L/W)_0$ is

$$\hat{\sigma}^{(\mathrm{L/W})} = \pi \left( \frac{1}{(L/W)_0 - 1} \right)^2 + \pi \left( \frac{1}{(L/W)_0 + 1} \right)^2 = 2\pi \frac{(L/W)_0^2 + 1}{[(L/W)_0^2 - 1]^2} \ , \tag{3.12}$$

and the cross section is obtained by multiplying $\hat{\sigma}$ with the squared length scale in the source plane,

$$\bar{\sigma}^{(\mathrm{L/W})} = \bar{\eta}_0^2 \hat{\sigma}^{(\mathrm{L/W})} \ , \tag{3.13}$$

where the bar on $\sigma$ is meant to indicate that the cross section was obtained from the analytical approximation rather than from the numerical cluster model.

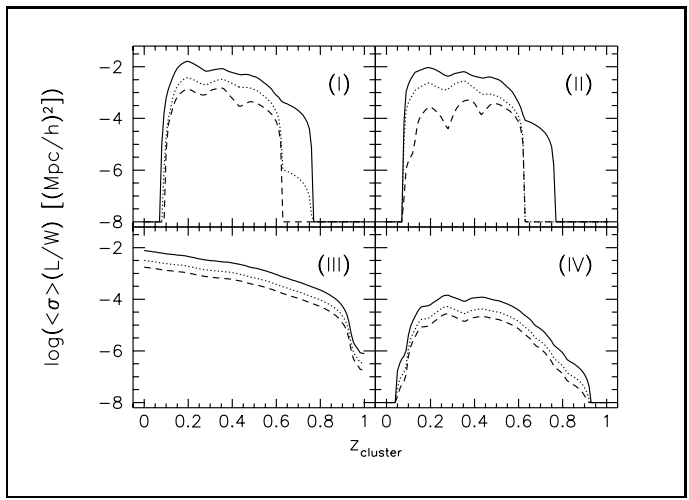

**Fig. 7.** Comparison of four sets of cross sections for the length-to-width ratio of arcs (solid, dotted, and dashed curves correspond to $L/W \geq 5, 7.5, 10$, as in the previous figures). Frames (I) and (II) show the cross sections for elliptical and circular sources, respectively, as already displayed in Fig. 6. Frame (III) shows the cross sections that result if each cluster model is replaced by a singular isothermal sphere with the velocity dispersion of the numerical model. Frame (IV) displays the cross sections resulting from modelling each cluster as a non-singular isothermal sphere with the velocity dispersion of the numerical model and the core radius taken to be half the numerical core radius

**3.2.2 Non-singular isothermal spheres.** For non-singular isothermal spheres, we introduce a dimensionless core radius $x_c$ to change the convergence of the lens to

$$\kappa = \frac{1}{2\sqrt{x^2 + x_c^2}} \qquad (3.14)$$

instead of $\kappa = 1/(2x)$ in the case of the singular isothermal sphere. The core radius $x_c$ is obtained by fitting the azimuthally averaged numerical surface-mass density profiles as described in Paper I, Sect. 3.3; see below. Then, the moduli of the eigenvalues are given by

$$\lambda_t = \left| 1 - \frac{m(x)}{x^2} \right| \;,\; \lambda_r = \left| 1 - 2\kappa(x) + \frac{m(x)}{x^2} \right| \;, \qquad (3.15)$$

where

$$m(x) \equiv \int_0^x dx' 2x' \kappa(x') \;. \qquad (3.16)$$

The length-to-width ratio of arcs is approximated by

$$\frac{L}{W} \simeq \max\left( \frac{\lambda_r}{\lambda_t}, \frac{\lambda_t}{\lambda_r} \right) \;, \qquad (3.17)$$

where we include radial arcs (with $\lambda_r < \lambda_t$) since they are also included in the numerical analysis. We find the region in the source plane where $(L/W) \geq (L/W)_0$ solving Eq. (3.17) numerically and obtain the cross section analogous to the singular case.

To adapt the analytic models to the numerical models, we choose the parameters $\sigma_v$ and $x_c$ obtained from the numerical models. The core radii of the numerical models were obtained from azimuthally averaging the surface mass density of the clusters. Therefore, whenever a cluster has significant substructure, the so-determined core radius will be larger than that of the dominant mass concentration. Thus, inserting the numerically determined cluster core radii into the non-singular isothermal sphere model might result in an unfair comparison, rendering the analytic approximations too weak lenses. To compensate for this, we divide the numerically determined core radii by two before inserting them into the non-singular isothermal models. Fig. 7 displays the results for the cross sections, and compares them with the numerically determined cross sections.

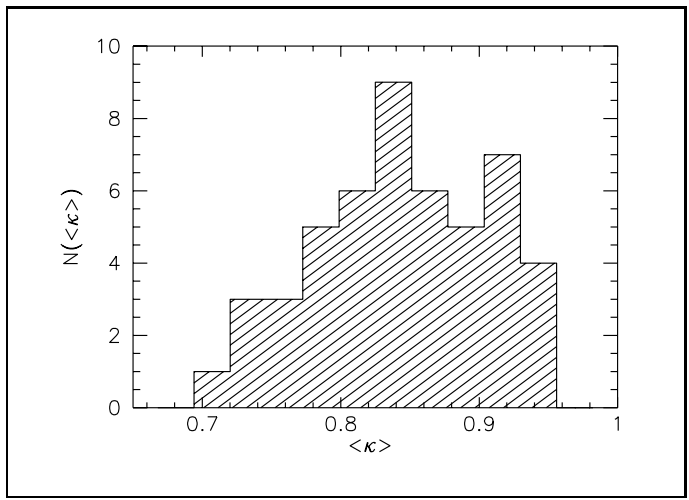

**Fig. 8.** Mean convergence $\langle\kappa\rangle_c$ within the critical curves of the numerical cluster models. The histogram shows that, on average, $\langle\kappa\rangle_c \simeq 0.85$, which reduces the mass discrepancy derived from comparing lensing and X-ray properties by a factor of $\simeq 1.2$, but does not resolve it

Fig. 7 shows that the cross sections derived from the non-singular isothermal sphere model (frame IV) is lower than the numerically determined cross sections (frames I and II) by roughly two orders of magnitude. The cross sections from the singular isothermal sphere model do not tend to zero for $z \to 0$ because this lens model remains critical for arbitrarily low redshifts due to their central density singularity. Thus, although their average amplitude at intermediate redshifts is about one order of magnitude lower than the numerically determined cross sections, they extend over a larger redshift range, which makes them as efficient for producing large arcs as are the numerical models.

### 3.3 Mass estimates

In cases where the convergence is rather low and the cross section is enhanced by shear, the averaged $\kappa$ within the critical curve $\langle\kappa\rangle_c$ of the cluster can be smaller than unity. In two recent papers (Miralda-Escudé & Babul 1994, Loeb & Mao 1994), it was argued, based on observations of three arc clusters, that these appear "too cold" for lensing when their density distribution inferred from their X-ray surface brightness distribution is compared to their lensing properties as derived from arcs observed within them. There, it was assumed that $\langle\kappa\rangle_c$ is unity, as is the case for spherically symmetric clusters. If now $\langle\kappa\rangle_c < 1$, the mass discrepancy implied by the above argument, i.e., that the mass required for lensing is about a factor of $2\ldots 3$ too high when compared to the mass required for their X-ray emission, is reduced. Therefore, we have determined $\langle\kappa\rangle_c$ for our numerical cluster models; the result is shown as a histogram in Fig. 8.

It can be read off from Fig. 8 that $\langle\kappa\rangle_c \simeq 0.85$ on average. Thus, the mass discrepancy is reduced by a factor of $\simeq 1.2$, but not resolved. However, in the papers cited above, it was assumed that the critical curve is a circle whose radius was taken to be the distance of dominant arcs from the cluster center. If the critical curves are elongated, and if arcs are preferentially located close to the far ends of the critical curves, this assumption would enclose regions where $\kappa$ is still smaller, which could further reduce $\langle\kappa\rangle_c$ and thus reduce the mass required for an explanation of the arcs.

## 4 Optical depths

Having determined cross sections, we turn to the computation of optical depths. The optical depth for lensing with property $Q_l$ is the probability that a given source is imaged

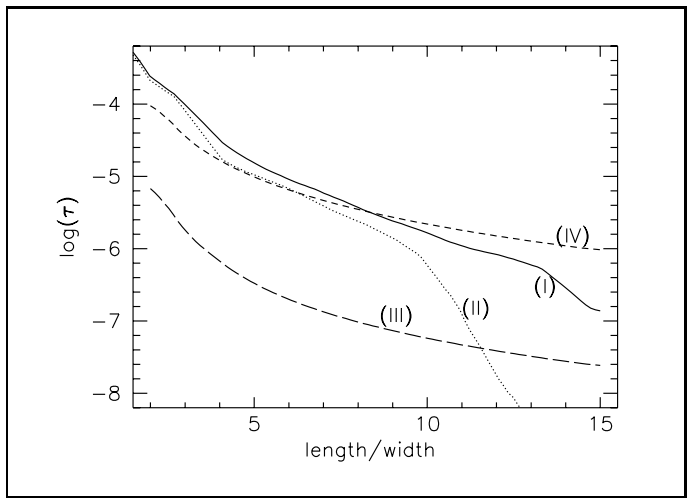

**Fig. 9.** Optical depth for the formation of arcs as a function of length-to-width ratio. Curve (I) (solid line) is obtained from the numerical cluster models with elliptical sources, curve (II) (dotted line) is for circular sources. Curve (III) (long-dashed line) shows the optical depth resulting if each cluster is replaced by a non-singular isothermal sphere, and curve (IV) (short-dashed line) displays the optical depth for singular isothermal spheres. The decline in curves (I) and (II) is caused by the finite source size and the finite grid resolution

with that property. Under the assumptions that (1) cross sections do not significantly overlap and (2) source positions are uncorrelated with lens positions, the optical depth $\tau$ is determined by adding up all cross sections attached to lenses between the observer and the source, and dividing the result by the area of the source plane,

$$\tau^{(l)}(z_\mathrm{s}) = \frac{1}{4\pi D_\mathrm{s}^2} \int_0^{z_\mathrm{s}} dz \left| \frac{dV(z)}{dz} \right| n_0 (1+z)^3 \sigma^{(l)}(z) \ . \tag{4.1}$$

Here, $n_0$ is the comoving number density of lenses, which is assumed to be constant, and the volume element $dV(z)$ is given by

$$dV(z) = \frac{c}{H_0} 4\pi D(z)^2 \frac{dz}{(1+z)^{5/2}} \tag{4.2}$$

for an Einstein-de Sitter universe. The assumption that the comoving density of lenses is constant is fulfilled in our case since clusters are neither created nor destroyed within $0 \le z \le 1$. We therefore have from Eqs.(4.1) and (4.2)

$$\tau^{(l)}(z_\mathrm{s}) = \frac{c}{H_0} n_0 \int_0^{z_\mathrm{s}} dz \sqrt{1+z} \left( \frac{D(z)}{D(z_\mathrm{s})} \right)^2 \sigma^{(l)}(z) \ . \tag{4.3}$$

We evaluate this integral inserting the cluster density within the large simulation volume,

$$n_0 = \frac{13}{(150 \text{ Mpc}/h)^3} \simeq 4 \times 10^{-6} \left( \frac{\text{Mpc}}{h} \right)^{-3} , \tag{4.4}$$

and the cross sections determined before. Fig.9 shows the results $\tau^{(\mathrm{L/W})}$ for $z_\mathrm{s} = 1$ as a function of $(L/W)$.

The solid and the dotted lines (curves I and II) in Fig.9 display the optical depths for the arc length-to-width ratio obtained from numerically determined cross sections for elliptical and circular sources, respectively. Curves (III) and (IV) (long- and short-dashed, respectively) are obtained from the analytic models, curve (III) from the non-singular and curve (IV) from the singular case. For all cases except case (III), the optical depth for long and thin arcs ranges between $10^{-5}$ and $10^{-6}$, while the optical depth for the non-singular approximation is lower by roughly two orders of magnitude. The optical depth from the numerical cluster models for either elliptical or circular sources (curves I and II, respectively) are similar for $(L/W) \lesssim 10$, while for larger length-to-width ratios the optical depth for elliptical sources is substantially larger. This indicates again that the source ellipticity is crucial for arcs with a large length-to-width ratio. For arcs of moderate $(L/W)$, the numerically determined optical depth for elliptical sources is about as large as for the singular isothermal cluster models, but it falls below for arcs with $(L/W) \gtrsim 13$. These results show that the numerical cluster models are significantly more efficient in producing large arcs than spherical analytic models would predict when adapted to the observable properties of the numerical models, and that the numerical models are roughly as efficient as singular isothermal models although they have extended cores. In Fig.10, we show the function

$$\frac{1}{\tau}\frac{d\tau}{dz} \equiv p_{\rm lens}(z) \qquad (4.5)$$

for $(L/W) \geq 7.5$. This function shows the fractional contribution of lenses within $dz$ of $z$ to the total optical depth and can thus be considered as a probability distribution for the redshift of lenses creating images of the specified length-to-width ratio.

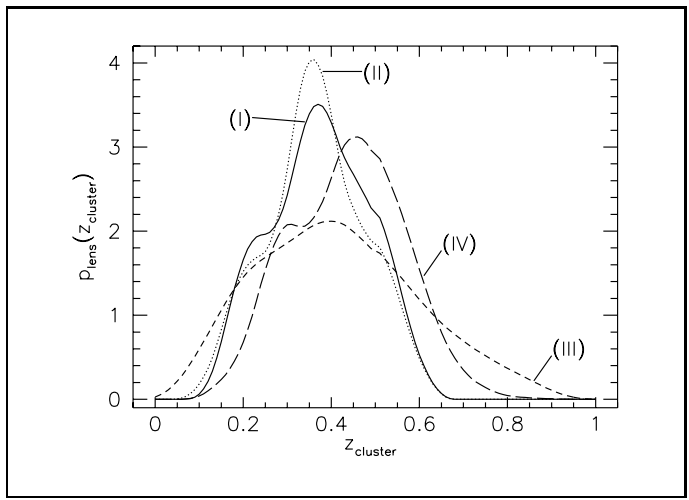

**Fig. 10.** The function $p_{\rm lens}(z_{\rm d})$ as defined in Eq.(4.5) for arcs with length-to-width ratio greater or equal 7.5; the curves show the fractional contribution of clusters within $dz_{\rm d}$ of $z_{\rm d}$ to the total optical depth and can thus be viewed as a probability distribution for cluster lenses. Curves (I) and (II) are for the numerical cluster models and elliptical or circular sources, respectively, while curves (III) and (IV) are for spherical cluster models, either singular or non-singular, respectively

The solid and the dotted curve (I and II, respectively) show $p_{\rm lens}(z)$ for elliptical and circular sources, respectively, while curves (III) and (IV) display $p_{\rm lens}(z)$ derived from the analytic models, either singular (curve III) or non-singular (curve IV). Curve (IV) is broadest, which reflects the fact that the cross section for singular isothermal spheres remains finite over the whole redshift range between observer and source due to their central density singularity, cf. frame (III) of Fig.7. This is not the case for the non-singular analytic model, but the maximum of $p_{\rm lens}(z)$ is slightly larger in this case than for the numerically determined distributions.

As a final application of the arc cross sections, we ask what the (cumulative) distribution of arc curvature radii and arc widths is within a sample of large arcs. This fraction is defined to be the optical depth for $(L/W) \geq (L/W)_0$ *and* either $R \geq R_0$ or $W \geq W_0$, which is determined in much the same way as the optical depth for only one of these arc properties, divided by the optical depth for $(L/W) \geq (L/W)_0$ alone. Fig.11a shows the distribution of arc curvature radii, Fig.11b the distribution of arc widths. The solid (dotted) curves in both figures are for elliptical (circular) sources, and the three curves per line type are for $(L/W)_0 \in \{5, 7.5, 10\}$.

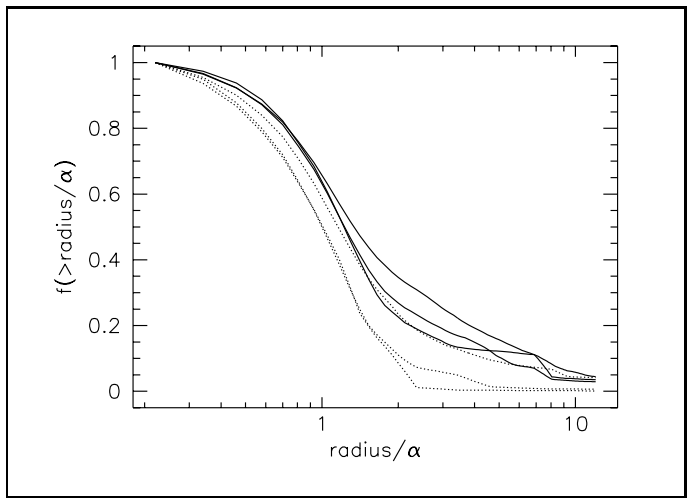

**Fig. 11a.** The fraction of arcs with curvature radius larger than specified among all arcs with length-to-width ratio $L/W \geq 5, 7.5, 10$. The solid curves are for elliptical, the dotted curves for circular sources. The curvature radii are normalized by the deflection angle $\alpha$ at the arc positions

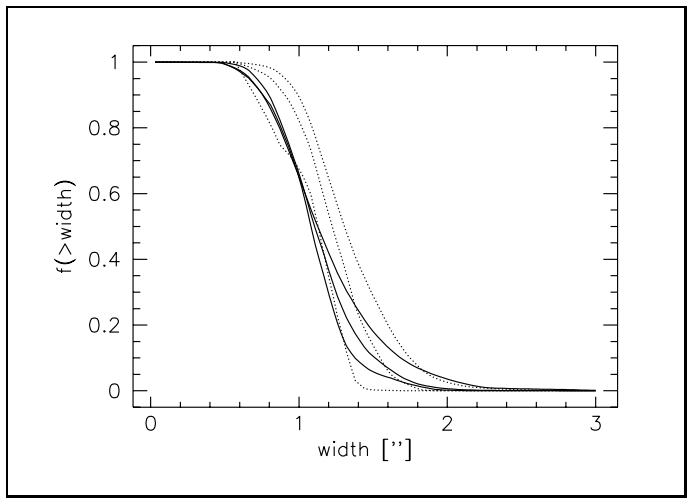

**Fig. 11b.** The fraction of arcs with curvature radius larger than specified among all arcs with length-to-width ratio $L/W \geq 5, 7.5, 10$. The solid curves are for elliptical, the dotted curves for circular sources

Following Miralda-Escudé (1993a), we normalize the arc curvature radii in Fig.11a by the deflection angle at the arc position. In case of a spherically symmetric lens model, this ratio would be unity for tangential arcs. The long tail in the radius distribution shows that there is a substantial fraction of 'straight arcs' formed, i.e., arcs with a very large curvature radius, and this is quite insensitive to the length-to-width ratio of the arc sample (cf. also Fig.3). This means that we expect that about 10% of all arcs should have a large curvature radius irrespective of their length-to-width ratio. There is a tendency that the curvature radii of arcs formed from circular sources are slightly smaller than those formed from elliptical sources.

The width distribution, displayed in Fig.11b, shows that most large arcs are about $1''$ wide, corresponding to the diameter of the circular sources chosen. This indicates that most large arcs are only mildly magnified or demagnified in their 'narrow' direction. For elliptical sources, the median width is tendentially lower than for circular sources except for arcs with a length-to-width ratio $\gtrsim 10$. The fraction of large arcs from elliptical sources with widths smaller than $1''$ is $\simeq 40\%$, quite independent of their length-to-width ratio, whereas for circular sources, the fraction of 'thin' arcs increases with increasing $(L/W)$.

# 5 Summary and Discussion

We constructed a sample of galaxy clusters within the CDM cosmogony and studied their capability to produce large arcs from background sources by their gravitational lensing effect. The clusters were selected from a large cosmological simulation box of side length 150 Mpc/$h$. The regions in the large box recognized as clusters were identified in the simulation volume at $z = 0$. We identified 13 regions in the large simulation whose total mass within a sphere of radius 5 Mpc/$h$ is larger than $10^{16}$ solar masses. Then, these regions were re-computed with higher resolution, we identified the center-of-mass in each of them, and cut out the particle distribution in boxes of comoving side length 5 Mpc/$h$ centered on the centers-of-mass. These boxes were considered as clusters. We computed their density distributions and the surface mass density fields projecting the boxes on each of their three independent sides, obtaining 39 surface density fields whose lensing capabilities were then studied, employing the techniques described in detail in Paper I. Three of these clusters were subcritical to lensing in all three projection directions, and two of these three were obtained from those regions in the large box which enclosed the lowest masses. We are therefore confident to have found all clusters in the simulation volume which can be considered lensing clusters, in other words, that our cluster sample is complete with respect to the lensing properties we seeked to investigate.

A total of 73395 images were classified. From this sample of images, we determined cross sections for the formation of arcs with several properties; the cross sections for arc length and arc length-to-width ratio were considered in closest detail. Having found cross sections, we computed the total optical depth for the formation of large arcs, the probability distributions of cluster lenses, and the distribution of arc curvature radii and arc widths within samples of large arcs. We also investigated the influence of intrinsic source ellipticity vs. intrinsically circular sources. The cross sections and optical depths found from the numerical cluster models were compared to analogous results derived from spherically symmetric analytic approximations to the numerical cluster models.
Our findings can be summarized as follows:

1. Cross sections for the formation of large arcs can have several local minima and maxima as functions of the cluster redshift. These result from merging processes in the clusters, which increase the surface mass density if the merger does not proceed along the line-of-sight, and rotation of the clusters, which can both increase and decrease the surface mass density. Thus, an individual cluster can be critical to lensing in two or more disconnected redshift intervals.
2. The cross sections for the formation of arcs with a length-to-width ratio $\gtrsim 10$ derived from the numerical cluster models are of order $10^{-3}$ (Mpc/$h$)$^2$ for intermediate cluster redshifts ($0.2 \lesssim z \lesssim 0.6$) and $z_\mathrm{s} = 1$.
3. The numerically modelled clusters produce cross sections for large arcs which are roughly two orders of magnitude larger than corresponding cross sections derived from non-singular isothermal cluster models with the same velocity dispersions and half the core radii of the numerical models.
4. The numerical cluster models are comparably efficient in producing large arcs as singular isothermal spheres with the same velocity dispersions, although they are non-singular.
5. The optical depths from the numerically modelled clusters and from the singular isothermal approximation are of order $3 \times 10^{-6}$ for arcs with a length-to-width ratio of $\simeq 10$, while it is about two orders of magnitude lower for the non-singular isothermal models.

6. Intrinsic source ellipticity is important in mainly two respects: first, the optical depth for arcs with $(L/W) \gtrsim 10$ is significantly larger for elliptical than for circular sources, and second, the large arcs formed from elliptical sources are narrower than those formed from circular sources.

When compared to the results of Wu & Hammer (1993, especially Figs.1&2 there), our results show that the numerical cluster models are as efficient for the formation of large arcs than their 'best' (i.e., most efficient) analytic model. This, however, had small cores and steep density profiles. In contrast, our numerical models have extended cores and fairly shallow density profiles, whose double-logarithmic slope is close to the isothermal value $-1$ (in projection) or slightly steeper. Moreover, a fraction of $\simeq 40\%$ of our large arcs from elliptical sources are narrow. We conclude that clusters neither have to have small cores nor steep density profiles to be efficient in producing arcs. We attribute these significant differences between the results from analytic and from numerical cluster models to the high abundance of significant substructures in clusters together with intrinsic deviations from spherical symmetry in the cluster components. These increase the shear, and can therefore render clusters critical to lensing whose surface mass density is markedly below its critical value. It is apparent from Figs.3 and 11a that circles traced by the arcs are approximately centered on the clusters' center of mass rather than on their substructures; this is because the substructures *alone* are uncritical to strong lensing.

This conclusion is also supported by the distribution of curvature radii of large arcs. Comparing our Fig.11a with Figs.7a and 7b of Miralda-Escudé (1993a), we see that among the large arcs produced by the numerical cluster models, the fraction of arcs with large curvature radii is significantly higher than found by Miralda-Escudé. Such 'straight arcs' occur when the caustic of the lens has either a lips or a beak-to-beak catastrophe. Arcs from sources close to a lips caustic are, however, located close to the cluster center and relatively unspectacular. This indicates that large arcs with large curvature radii are to be expected preferentially from sources at beak-to-beak caustics, which are formed when there are two or more mass centers in the lens.

We have also investigated whether there is evidence for projection effects. As proposed by Miralda-Escudé (1993a), possibly such clusters are most efficient for forming large arcs which are elongated along the line-of-sight towards the observer. For each cluster model used in this paper, we have compared the orientation of the principle-axis system of the cluster's inertial tensor relative to the coordinate axes of the large-scale simulation box, which are identical to the lines-of-sight for the three directions of projection. We have found no significant correlation between the lensing capability of the clusters (expressed by the sizes of their cross sections) and alignments between the two frames of reference. There are cases of large cross sections for which the alignment is very close, but the opposite also occurs. We conclude that cross sections can be large because of projection effects which enhance the surface-mass density, but that the larger shear occurring when substructured clusters are seen "from the side" can also cause large cross sections. Thus, when clusters are elongated along the line-of-sight, the cross section can be large because of the enhanced convergence, and when clusters are seen "from the side", the cross section can be enlarged by higher shear. Due to the enhanced shear, the mean convergence enclosed by the critical curves $\langle \kappa \rangle_\mathrm{c}$ is smaller than unity in the numerical cluster models (on average, $\langle\kappa\rangle_c \simeq 0.85$). As argued above, this reduces the mass discrepancy derived from a comparison between X-ray and lensing properties of clusters which was recently discussed by Miralda-Escudé & Babul (1994) and Loeb & Mao (1994), but does not resolve it.

Finally, substructure is frequent in clusters when the density parameter of the Universe $\Omega_0$ is close to unity (e.g., Richstone et al. 1992, Bartelmann et al. 1994, Evrard et al. 1993). If substructure is as important for lensing statistics as implied by the numerical simulations presented here, and if arc searches continue to be as successful as they have been until now, substructure is required to understand the observed arc statistics. This provides a further argument for $\Omega_0$ to be close to unity. If $\Omega_0$ were much smaller, e.g., $\Omega_0 = 0.2$, then clusters form early and have sufficient time to relax and form spherically symmetric mass distributions, and these are significantly less efficient in producing arcs than unrelaxed, substructured clusters.

*Acknowledgements.* We are grateful to Peter Schneider and Martin Hähnelt for many discussions and for their stimulating comments.